\documentclass[useAMS,usenatbib,usegraphicx]{mn2e}

\usepackage{amsmath,amssymb,amsbsy,amsfonts}
\usepackage{amssymb}
\usepackage[british]{babel}
\usepackage{mdwmath}
\usepackage{color}
\usepackage{url}

\title[A natural approach to extended Newtonian gravity]{A 
       natural approach to extended Newtonian gravity: tests
and predictions across astrophysical scales}
\author[S. Mendoza, et al.]
       {S. Mendoza, X. Hernandez, J.C. Hidalgo \& T. Bernal \\
           Instituto de Astronom\'{\i}a, Universidad Nacional
           Aut\'onoma de M\'exico, AP 70-264, Distrito Federal 04510,
           M\'exico
           }

\volume{VER10}
\pagerange{\pageref{firstpage}--\pageref{lastpage}}

\begin{document}

\maketitle
\label{firstpage}
\begin{abstract}
  In the pursuit of a general formulation for a modified gravitational
theory at the non-relativistic level and as an alternative to the dark
matter hypothesis, we construct a model valid over a wide variety of
astrophysical scales.  Through the inclusion of Milgrom's acceleration
constant into a gravitational theory, we show that very general formulas
can be constructed for the acceleration felt by a particle.  Dimensional
analysis shows that this inclusion naturally leads to the appearance
of a mass-length scale in gravity, breaking its scale invariance.
A particular form of the modified gravitational force is constructed
and tested for consistency with observations over a wide range of
astrophysical environments, from solar system to extragalactic scales.
We show that  over any limited range of physical parameters, which define
a specific class of astrophysical objects, the dispersion velocity of a
system must be a power law of its mass and size.  These powers appear
linked together through a natural constraint relation of the theory.
This yields a generalised gravitational equilibrium relation valid for
all astrophysical systems.  A general scheme for treating spherical
symmetrical density distributions is presented, which in particular
shows that the fundamental plane of elliptical galaxies, the Newtonian
virial equilibrium, the Tully-Fisher and the Faber-Jackson relations,
as well as the scalings observed in local dwarf spheroidal galaxies,
are nothing but particular cases of that relation when applied to the
appropriate mass-length scales.  We discuss the implications of this
approach for a modified theory of gravity and emphasise the advantages
of working with the force, instead of altering Newton's second law of
motion, in the formulation of a gravitational theory.
\end{abstract}


\section{Introduction}
\label{intro}

  The dynamical mass to light ratios derived for spiral galaxies are usually
much greater than expected for their stellar components. This is often
interpreted as indicating the gravitational dominance of hypothetical
dark matter.  Alternatively, one could argue that the discrepancy between
dynamical mass and baryonic mass is telling us that the Newtonian law
of gravity is not the one governing the dynamics. In particular, the
MOdified Newtonian Dynamics (MOND) proposed by \citet{milgrom83a} has been
proven to be successful in explaining how galaxies rotate, without any
dark matter \citep[see e.g.][for a review]{sanders02}.

  Recently, the range of astrophysical problems treated under the MOND
hypothesis has increased significantly.  Abundant recent publications
on velocity dispersion measurements for stars in the local dwarf
spheroidal galaxies, the extended and flat rotation curves of spiral
galaxies, the large dispersion velocities of galaxies in clusters,
the gravitational lensing due to massive clusters of galaxies, and
even the cosmologically inferred matter content for the universe,
have been successfully modelled under MOND.  These, not as indirect
evidence for the existence of a dominant dark matter component, but as
direct evidence for the failure of the current Newtonian and general
relativistic theories of gravity, in the large scale or low acceleration
regimes relevant for the above \citep[see ][for recent examples]{
sanders07,nipoti07,famaey07,gentile07,tiret07,sanchez08,bekenstein04,capozziello07,sobouti07,mendoza07}.
MOND has proved successful on many astrophysical situations, though
difficult on others  \citep[see e.g.][for a good review on these
points]{milgrom08-paradigm,milgrom09,bekenstein06,zhao05}.

  The key feature of Milgrom's Modified Newtonian Dynamics is
the introduction of a fundamental acceleration scale \( a_0 =
1.2 \times 10^{-10} \mathrm{m}\,\mathrm{s}^{-2} \) \citep[see
e.g.][]{milgrom08-paradigm} into gravitation.  The introduction of \(
a_0 \) can alternatively be regarded as grounded upon direct empirical
evidence, as the observed dynamics of large spiral discs attest.
Additionally, non-relativistic gravity due to a point mass \( M \),
results in a force on any test particle such that it is pulled  towards \(
M \) with an acceleration \( a \).  Fundamentally, Newton's constant of
gravity \( G \) completes the description of the problem and so, by means
of Buckingham's theorem of dimensional analysis \citep[cf.][]{sedov}, the
absolute value of the attractive acceleration felt by the test particle
located at a distance \( r \) from the point mass \( M \) is given by

\begin{equation}
  a = a_0 \, f(x),
\label{eq001}
\end{equation}

\noindent where

\begin{gather}
  x := l_\text{M} / r,
						\label{eq-dimensionless} 
\intertext{with,}
  l_\text{M}  :=  \left( \frac{ G M }{ a_0 }  \right)^{1/2}.
 						\label{eq-lm}
\end{gather}

\noindent The acceleration expressed in equation~\eqref{eq001} converges
to Newton's gravitational acceleration when the function \( f(x) = x^2 \)
and to MOND's acceleration when \( f(x) = x \).  These two examples of
functions \( f \) represent the gravitational approach to an extended
non-relativistic theory of gravity.  The main problem is how to find a
function \( f(x) \) which, for the appropriate limits, converges to the
Newtonian and MONDian regimes.

  \citet{hernandez10} showed that the \citet{bekenstein04} function \(
f(x) = x + x^2 \) serves quite well when applied to dwarf spheroidal
(dSph) galaxies and to the rotation curves at large radii of spiral
galaxies, but is inconsistent with measured limits on departures from
Newtonian gravity at solar system scales.  Also, \citet{binney-famaey}
showed that this particular form of the function \( f(x) \) does
not work well when applied to our Galaxy.  Despite the fact that this
prescription has the corresponding limits as expected (\( f(x) \to x \)
when \( x \ll 1 \) and \( f(x) \to x^2 \) when \( x \gg 1 \)), a more
general function must be constructed.

  Note that pure dimensional analysis, with the introduction of
an acceleration scale \( a_0 \), determines exactly the
dimensional form that the acceleration must have.   In very general terms,
it also shows that the introduction of \( a_0 \) means that gravity
has a characteristic mass-length scale \( l_\text{M} \) which makes
possible the construction of equation~\eqref{eq001}.  With all these,
the acceleration turns out to be a function of the variable \( x \) only,
which as we will show later, gives a robust way of working with an extended
theory of gravity at the non-relativistic level.

 It is important to note that Milgrom has already introduced the length
\( l_\text{M} \) \citep[see e.g.][where it appears as a transition
radius]{milgrom83a,milgrom83b,milgrom86,milgrom08,milgrom08-paradigm}.
In these studies it is shown that this mass-length scale serves as a
transition point where the MONDian regime passes to the Newtonian one.
\citet{milgrom08} stressed the points that a mass distribution whose
length is much greater than its associated mass-length \( l_\text{M} \)
is in the MONDian regime (since \( x \ll 1 \)) and a mass distribution
whose length is much smaller than its mass-length scale is in the
Newtonian regime (since \( x \gg 1 \)). The case \( x = 1 \) can roughly
be thought of as the point where the transition from the Newtonian to
the MONDian regimes occurs.

  We now show that there is a connection between this approach and the one
commonly used in the implementation of the MOND theory.  The physical form
of MOND is given by \citep[see e.g.][]{bekenstein-milgrom84,milgrom01}
an Aquadratic Lagrangian (AQUAL) and so, its variation reproduces the
equation of motion. For relevant symmetries in the problem, this approach
gives the important result that the absolute value of the acceleration
felt by a test particle in the presence of a point mass is given by

\begin{equation}
  a \, \, \mu\left( a / a_0 \right) = | \nabla \phi_\text{N} | 
     = \frac{ G M }{ r^2 }.
\label{eq00a}
\end{equation}

\noindent In this equation, the Newtonian scalar potential is represented
by \( \phi_\text{N} \) and the interpolation function \( \mu(a/a_0) \) is
such that \( \mu(a/a_0) = 1 \) in the Newtonian limit, which corresponds
to \( a \gg a_0 \) and \( \mu(a/a_0) = a/a_0 \) in the MONDian regime,
with \( a \ll a_0 \). Equating the acceleration in relation~\eqref{eq00a}
with that of equation~\eqref{eq001}, it follows that

\begin{equation}
  \mu(a/a_0) = \frac{x^2}{f(x)},
\label{eq002}
\end{equation}

\noindent which implicitly shows that the MOND formalism can be
equivalently expressed through the modification of the gravitational
force~\eqref{eq001}.  We note here that the MOND formulation \eqref{eq00a}
refers to a modification of the dynamical sector of the theory, whereas
equation~\eqref{eq001} is completely based on the modification of the
gravitational force. Both are operationally equivalent formulations.
The MOND formulation has always been tackled through dynamical
modifications.   However, we show in this article that there are many
advantages when choosing the modification in the gravitational sector.
Therefore we only use the constant \( a_0 \) for consistency with
the dynamical modifications. It is important to emphasise that in
the gravitational modifications it is more natural to frame the
problem in terms of the mass-length scale \( l_\text{M} \) defined
in equation~\eqref{eq-lm}.  Furthermore, it is the use of dimensional
analysis which tells us the very important fact that the dimensionless
force \( f(x) \) in equation~\eqref{eq001} only depends on the ratio \(
l_\text{M} / r \).

  The article is organised as follows. Section~\ref{model} 
introduces a particular form of the function \( f(x) \).
This is used in the subsequent sections for applications in
different astrophysical environments, from solar system to
galaxy cluster scales.  Finally, in Section~\ref{discussion} we discuss
the advantages of such a general function \( f(x) \).

\section{The force model}
\label{model}

  The dimensionless gravitational force \( f(x) \) in equation~\eqref{eq001} 
felt by a given test particle must be analytic, 
and as such it can be written as

\begin{equation}
  f(x) = \sum_{n=-\infty}^{n=\infty} \, c_n x^n.
\label{eq003}
\end{equation}

\noindent  We now show how to obtain a reasonable \( f(x) \) by simple
exploration of the Newtonian and MONDian regimes.  First of all, notice
that in the Newtonian and deep MOND limits, the function \( f(x) \) is
such that \( f(x) = c_\text{N} \, x^2 \) and \( f(x) = c_\text{M} \, x \)
respectively, with \( c_\text{N} = c_\text{M} = 1 \).  We now focus on
the Newtonian \( x \gg 1 \) regime and explore an expansion about that 
limit of the form

\begin{equation}
  \begin{split}
  \left( \frac{ a }{ a_0 } \right) &= 
    \left( \frac{ l_\text{M} }{ r } \right)^2 + 
    \left( \frac{ l_\text{M} }{ r } \right) + 
    \left( \frac{ l_\text{M} }{ r } \right)^0 +  
    \left( \frac{ l_\text{M} }{ r } \right)^{-1} +  
    \ldots, \\
  &= \left( \frac{ l_\text{M} }{ r } \right)^2 \left\{
    1 + 
    \left( \frac{ l_\text{M} }{ r } \right)^{-1} + 
    \left( \frac{ l_\text{M} }{ r } \right)^{-2} + 
    \left( \frac{ l_\text{M} }{ r } \right)^{-3} + 
    \ldots
    \right\}, \\
  &= x^2 \left( 1 + x^{-1} + x^{-2} + x^{-3} + \ldots \right).
  \end{split}
\label{eq01}
\end{equation}

\noindent The limit \( x \to \infty \) gives the Newtonian
acceleration, so taking into account all the terms of the geometric
series for \( x > 1 \) then,

\begin{equation}
  \left( \frac{ a }{ a_0 } \right) =  x^2 \left( 1 + 
    \frac{ 1 }{ x - 1 } \right)  = \frac{ x^3 }{ x - 1 }.
\label{eq02}
\end{equation}

  We now put special emphasis on the MONDian \( x \ll 1 \) regime and explore 
the corresponding expansion, given by

\begin{equation}
  \begin{split}
  \left( \frac{ a }{ a_0 } \right) &= 
    \left( \frac{ l_\text{M} }{ r } \right) + 
    \left( \frac{ l_\text{M} }{ r } \right)^2 + 
    \left( \frac{ l_\text{M} }{ r } \right)^3 +  
    \left( \frac{ l_\text{M} }{ r } \right)^{4} +  
    \ldots, \\
  &= \left( \frac{ l_\text{M} }{ r } \right) \left\{
    1 + 
    \left( \frac{ l_\text{M} }{ r } \right) + 
    \left( \frac{ l_\text{M} }{ r } \right)^{2} + 
    \left( \frac{ l_\text{M} }{ r } \right)^{3} + 
    \ldots 
    \right\}, \\
  &= x \left( 1 + x + x^{2} + x^{3} + \ldots \right).
  \end{split}
\label{eq03}
\end{equation}

\noindent Note that the deep MOND regime is obtained in the limit 
\( x \to 0 \), and so the geometric series of equation
\eqref{eq03} for \( x < 1 \) gives

\begin{equation}
\left( \frac{a}{a_0} \right) = \frac{ x }{ 1 - x }.
\label{eq04}
\end{equation}

  Equation~\eqref{eq01} can be thought of as the series for the negative
powers of relation~\eqref{eq003} and equation~\eqref{eq03} as the one
for the positive powers of the same relation.  The interesting thing to
note is that both of them can be analytically continued for all values
of \( x \). For the limit cases, the minus sign
on the denominator of both equations~\eqref{eq02} and~\eqref{eq04}
can be changed for a positive sign.

  Since we are interested in the complete analytic series let us propose
a general acceleration formula given by the addition or substraction of
equations \eqref{eq02} and \eqref{eq04} as follows:

\begin{equation}
\left( \frac{a}{a_0} \right)_\pm = \frac{ x \pm x^3 }{ 1 \pm x }.
\label{eq05}
\end{equation}

\noindent Note that this last equation tends to the Newtonian acceleration
regime when \( x \to \infty \) and to the MONDian acceleration
limit when \( x \to 0 \).  In fact, due to the symmetry of the
numerator and denominator of equation \eqref{eq05}, a more general
relation can be postulated:

\begin{equation}
  \left( \frac{a}{a_0} \right)_\pm = x \, \frac{ 1 \pm x^{n+1} }{ 1 \pm x^n
    }.
\label{eq06}
\end{equation}

\noindent  This satisfies the Newtonian and MONDian acceleration limits
for \( x \to \infty,\ 0 \) respectively.  Note also that the case \( n
= 1 \) with a minus sign is the same as two times the case \( n = 0 \)
with a plus sign, and both correspond to the \emph{Bekenstein ground state}
acceleration formula \citep{bekenstein04}.  This has proved to be useful
for the dynamical modelling of dSph galaxies \citep{hernandez10}, but not 
for our own Galaxy \citep{binney-famaey}.

  The acceleration function \eqref{eq06} has no singularities, since
according to l'H\^{o}pital's rule, \( a / a_0 \to (n+1)/n \) as \( x \to 1 \).
In fact, to see this directly, notice that for the minus sign it follows
from equation \eqref{eq06} that

\begin{equation}
  \begin{split}
  \left( \frac{a}{a_0} \right)_- &= x \, \frac{ ( 1 - x) ( 1 + x + x^2 + x^3 +
    \ldots + x^{n} ) }{ ( 1 - x ) ( 1 + x + x^2 + \ldots + x^{n-1} ) }, \\
  &= x \, \frac{ ( 1 + x + x^2 + x^3 +
    \ldots + x^{n} ) }{ ( 1 + x + x^2 + \ldots + x^{n-1} ) }.
  \end{split}
\label{eq06a}
\end{equation}

  For further applications we note that the right hand side of
equation~\eqref{eq06} with a minus sign can be Taylor expanded as follows:

\begin{gather}
  \left( \frac{a}{a_0}  \right)_- = x + x^{n+1} - x^{n+2} + x^{2n+1}
    - x^{2n+2} + \ldots, 				\label{a01} \\
\intertext{ for $x < 1$, and}
  \left( \frac{a}{a_0}  \right)_- = x^2 - x^{1-n} + x^{2-n} - 
    x^{1-2n} + x^{2-2n} + \ldots,			\label{a02} \\
\intertext{ for $x > 1$. Choosing the positive sign we obtain:}
  \left( \frac{a}{a_0}  \right)_+ = x - x^{n+1} + x^{n+2} + x^{2n+1}
    - x^{2n+2} + \ldots,				\label{a03} \\
\intertext{for $x < 1$, and}
  \left( \frac{a}{a_0}  \right)_+ = x^2 + x^{1-n} - x^{2-n} - 
    x^{1-2n} + x^{2-2n} + \ldots, 			\label{a04} 
\end{gather}

\noindent for \( x > 1 \). This shows that the Newtonian and MONDian
regimes are reached in the correct limit regardless of the value of \(
n \).

  The extreme limiting case of \( n \to \infty \) corresponds to the
function

\begin{equation}
  \left( \frac{ a }{ a_0 } \right)_\text{e}  = 
  \begin{cases}
    x, \qquad \text{  for }  0 \leq x \leq 1  \quad \ \text{(MONDian
    regime).} \\
    x^2, \qquad \text{for  } x \geq 1  \qquad \quad \text{(Newtonian
    regime).}
  \end{cases}
\label{eq07}
\end{equation}

\noindent This acceleration formula is of no use due to the discontinuity
on the first derivative at \( x = 1 \), but serves as a reference
to understand that the real acceleration must smoothly pass from the
Newtonian to the MONDian regime.  Also, as noted by \citet{milgrom08},
the point where \( x = 1 \) represents approximately the transition
from the Newtonian to the MONDian regimes.  In the proposed model,
this is strictly valid in the extreme case, with \( n \to \infty \).
This point is relevant, as any function moving away from the value \(
a = a_0 \) at \( x = 1 \) seems to have a better chance at modelling a
more real astrophysical situation, since it will smoothly transit from
the MONDian regime to the Newtonian one.

\begin{figure}
\begin{center}
  \includegraphics[scale=0.7]{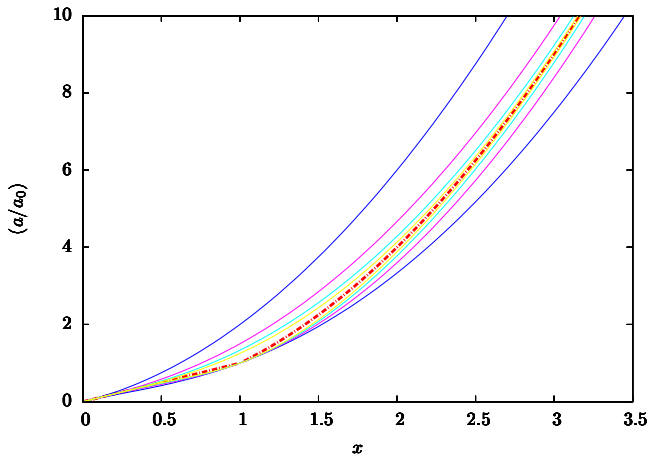}
\end{center}
\caption[Acceleration]{The figure shows the acceleration function \( a \) in
units of Milgrom's constant \( a_0 \) as a function of the parameter \( x
\). The thick dash-dot curve is the extreme limiting value \( n \to \infty \),
i.e. \( a / a_0 = x \) for \( x \leq 1 \) and \( a / a_0 = x^2 \) for \( x
\geq 1 \).  The curves above and below this extreme acceleration line
represent values of \( n = 4,\ 3,\ 2,\ 1, \) for the minus and plus
signs of equation \eqref{eq06} respectively.  The extreme limiting 
curve has a kink at \( x = 1 \) and is of no physical interest due to the
undefined derivative at that point.
}
\label{fig01}
\end{figure}

  Figure \ref{fig01} shows a plot of \( a / a_0 \) as a function
of \( x \) for various values of \( n \) and choice of signs in
equation~\eqref{eq06}.  From this figure it is seen that the curves
with \( n \sim 3 - 4 \) are very close to the extreme limiting case,
but preserve a good soft transition region between the MONDian and
Newtonian regimes.  However, it is through fits with observations that
an optimal real number \( n \) is to be calibrated.  In fact, the fit
with observations must also give us a way to decide between the plus or
minus sign in equation \eqref{eq06}.

%%%%%%%%%%%%%%%%%%%%%%F I G U R E%%%%%%%%%%%%%%%%%%%%%%%%%%%%%%%%%%%%%%%%%
%%%%%%%%%%%%%%%%%%%%%%F I G U R E%%%%%%%%%%%%%%%%%%%%%%%%%%%%%%%%%%%%%%%%%

  In order to fix the parameter \( n \) and the choice of sign in
equation \eqref{eq06}, we construct different corresponding
MOND interpolation functions \( \mu(a/a_0) \) for this acceleration
formula and compare them with the best model for our Galaxy presented
by \citet{binney-famaey}.   To do that, we must substitute
equations~\eqref{eq06} and~\eqref{eq001} into \eqref{eq002}, yielding

\begin{equation}
  \mu(a/a_0) =  x  \, \frac{ 1 \pm x^n }{ 1 \pm x^{n+1} }.
\label{eq08}
\end{equation}

\begin{figure}
\begin{center}
  \includegraphics[scale=0.7]{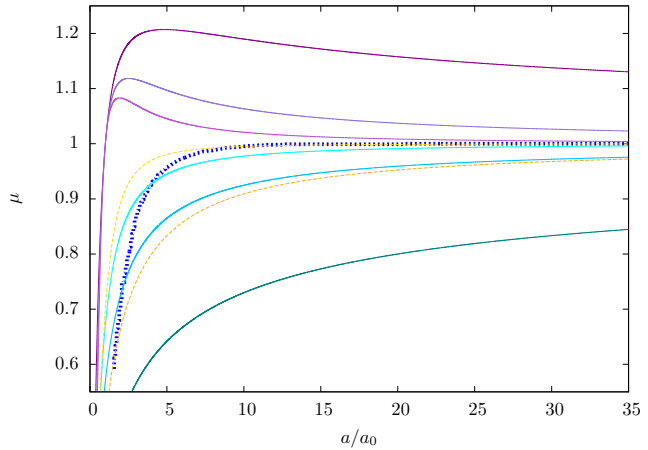}
\end{center}
\caption[Interpolation]{The figure shows MOND's interpolation
function \( \mu(a/a_0) \) for our Galaxy (dotted curve) as inferred by
\citet{binney-famaey}. From bottom to top (not counting the dotted curve)
the plots represent the following models: (1)~\citeauthor{bekenstein04}
ground state acceleration model \citep{bekenstein04}, which
corresponds to our equation \eqref{eq06} with \( n = 1 \) and a
minus sign. (2)~The interpolation formula \( \mu(\chi) = \chi / (
1 + \chi ) \) \citep[][dashed]{binney-famaey}. (3)~\( n=2 \) for a
minus sign. (4)~\( n=3 \) for a minus sign.  (5)~The interpolation
formula \( \mu(\chi) = \chi / \left( 1 + \chi^2 \right)^{1/2} \)
\citep[][dashed]{milgrom83a}. (6)~\( n = 3 \), plus sign. (7)~\( n=2 \),
plus sign. (8)~\( n=1 \), plus sign.  }
\label{fig02}
\end{figure}

\noindent The function \( x(a/a_0) \) that appears in equation
\eqref{eq08} is obtained by solving numerically equation \eqref{eq06}
for a fixed value of \( n \).  Figure \ref{fig02} shows that the best
fit to the optimal \( \mu(a/a_0) \) obtained by \citet{binney-famaey}
for our Galaxy, is reproduced with the minus sign and with \( n = 3.13
\approx 3 \). The effective gravitational acceleration formula is hence 
chosen as

\begin{equation}
  \left( \frac{a}{a_0} \right) = f(x) = x \, \frac{ 1 - x^4}{ 1 - x^3 } = 
    x \, \frac{ 1 + x + x^2 + x^3 }{ 1 + x + x^2 }.
\label{star}
\end{equation}

%%%%%%%%%%%%%%%%%%%%%%F I G U R E%%%%%%%%%%%%%%%%%%%%%%%%%%%%%%%%%%%%%%%%%
%%%%%%%%%%%%%%%%%%%%%%F I G U R E%%%%%%%%%%%%%%%%%%%%%%%%%%%%%%%%%%%%%%%%%

\noindent  Given observational errors, and uncertainties in the mass to
light ratios and their radial variability in our galaxy, the constrains
of Figure~\ref{fig02} are subject to considerable uncertainties.
As such, we intend to show mearly that functions of the proposed family
are clearly consistent with available estimates of \( \mu(a/a_0) \),
in this rather uncertain parameter range.  In the following section
we show this particular \( f(x) \) as consistent with the much more
stringent constraints available and the quasi--Newtonian scale of the
solar system.  Also, simple explanations for the observed structural
relations of elliptical and dwarf spheroidal galaxies will be shown
to appear naturally in Section~\ref{equilibrium}.  Note that for this
particular case, the function \( \mu(a/a_0) \) has an analytic solution,
since the function \( x(a/a_0) \) from equation \eqref{eq06} with \(
n=3 \) is the root of a fourth order polynomial in \( x \).  However,
due to its complicated form, we omit it here. It is interesting that
for this value of \( n \), the expansion about $x \gg 1$ begins with
the Newtonian term, and then skips the following two terms according to
equation~\eqref{a02}. This guarantees that dynamics will remain extremely
close to Newtonian for a large range of values of $x>1$ and sheds light
on the extended Newtonian character of the force of gravity. Similarly,
for \( x \ll 1 \), the leading term of equation~\eqref{a01} gives
the deep MOND regime \( x \) with the following one being \( x^4 \).
The absence of the \( x^2 \) and \( x^3 \) terms implies that physics
close to the deep MOND regime will not present any strong variations
for a considerable range of values in \( x < 1 \).

  We also note that the acceleration function~\eqref{eq001}  is such
that Newton's theorems, i.e. the acceleration field at distance $r$
from the centre of a spherical system depends only on the total mass
$M(r) $ interior to $r$, while external shells result in no force,
are valid for any analytic function~\eqref{eq003} which depends on the
parameter \( x \) only.  In order to see this, suppose that \( a \propto
x^p \) with \( p \) an integer number.  Assume that the test particle
is placed at position \( r \) inside a spherically symmetric shell.
If we trace a cone with solid angle $\delta \Omega$ and vertex at
the test particle, the shell is intersected at two opposite points
$\boldsymbol{r}_1$ and $\boldsymbol{r}_2$. The masses $\delta M_1$ and
$\delta M_2$ contained within the solid angle $\delta \Omega$ at these
points keep the proportion \( \delta M_2 / \delta M_1 = \left(r_2 / r_1
\right)^2 \).  This relation means that $({\delta M_1}/{r_1^2})^{p/2}
= ({\delta M_2}/{r_2^2})^{p/2} $, and so \( \delta  x^p(r_1) = \delta
x^p(r_2) \).  In other words, the acceleration exerted by the outer shell
at position \( r \) cancels out.  Since we can do this for any integer
\( p \), it follows that any analytical function of $x$ (cf.
equation~\eqref{eq003})
guarantees Newton's theorems. This is an encouraging
property of the acceleration~\eqref{eq001}.

  We now extend equation~\eqref{eq001} to a vector form.  As
noted above, the AQUAL formulation is satisfied by our model and as such
its field equation (Poisson's generalisation formula) is given by
\citep{bekenstein-milgrom84}

\begin{equation}
  \nabla \cdot \left[ \mu(a/a_0) \, \nabla \phi \right] = 4 \pi G \rho =
  \nabla^2 \phi_\text{N},
\label{e01}
\end{equation}

\noindent where the scalar potential \( \phi \) satisfies the condition
\( \boldsymbol{a} = - \nabla \phi \).   For systems with a high
degree of symmetry, \citet{bekenstein-milgrom84} showed that

\begin{equation}
  \left( \frac{ \boldsymbol{a} }{ a_0 } \right) \mu\left(a/a_0\right) = 
    - \frac{ 1 }{ a_0 } \nabla \phi_\text{N}  = - \frac{ G }{ a_0 } \int
      \frac{  \left( \boldsymbol{r} - 
      \boldsymbol{r}' \right) }{ 
      | \boldsymbol{r} - \boldsymbol{r}'|^3 } \, \rho(\boldsymbol{r}') 
      \, \mathrm{d} V',
\label{e02}
\end{equation}

\noindent where we can now express  the interpolation function \(
\mu \) through expression~\eqref{eq002}, and \( \rho(\boldsymbol{r}) \)
represents the matter density at the coordinate point \( \boldsymbol{r}
\). On the other hand, the vector form of equation~\eqref{eq001} must
necessarily be of the form

\begin{equation}
  \boldsymbol{a} = a_0 f(x) \boldsymbol{e}_a,
\label{e03}
\end{equation}

\noindent  where the unit vector \( \boldsymbol{e}_a \) points in the
direction of the acceleration \( \boldsymbol{a} \).  Substitution of
equation~\eqref{e03} into~\eqref{e02} with help of relation~\eqref{eq002} 
yields

\begin{equation}
  x^2 \boldsymbol{e}_a = - \frac{ G }{ a_0 } \int
      \frac{  \left( \boldsymbol{r} - 
      \boldsymbol{r}' \right) }{ 
      | \boldsymbol{r} - \boldsymbol{r}'|^3 }  \, \rho(\boldsymbol{r}')
      \mathrm{d} V'.
\label{e04}
\end{equation}

\noindent This description generalises the previous relations in the sense
that a point mass \( M \) can be directly substituted for \( M(r) \) in all
relevant equations.

  The magnitude of the vectorial equation~\eqref{e04} is the formula to
calculate the variable \( x \) for a given mass distribution density \(
\rho(\boldsymbol{r}) \).  As an example, for a point mass \( M \) the
density is given by \( \rho(\boldsymbol{r}) = M \delta(\boldsymbol{r})
\), where \( \delta \) represents Dirac's delta function. Consequently,
the variable \( x \) is given by equation~\eqref{eq-dimensionless}
as expected.  For the case of a spherically symmetric distribution of
matter, the density depends only on the radial coordinate \( r \).
This means that \citep{binney08} \( x^2 \boldsymbol{e}_a = - G M(r)
\boldsymbol{e}_r / r^2 \), where \( M(r) \) is the mass contained within
the radius \( r \) and \( \boldsymbol{e}_r \) is a unit vector in the
direction of the radial coordinate.  
With this, it follows that the variable \( x \) depends only on the mass
contained within the radius \( r \).  Since the general acceleration
function \( f(x) \) is analytic (cf. equation~\eqref{eq003}), then it
is also clear from this point of view that Newton's theorems are valid
for the class of models presented in this article.  

  Recently \citet{milgrom10} has developed a quasi-linear formulation
of MOND, which in particular can by applied to our spherical symmetric
case \citep[see also][]{zhao10}.  We now show that this is equivalent
to our formulation.  In this theory, a general gravitational potential
\( \Phi := \phi_\text{N} + \varphi \) is proposed, where \( \varphi \)
satisfies the equation

\begin{equation}
  \nabla^2 \varphi = \nabla \cdot \left[ \nu\left( |\nabla \phi_\text{N}|
    / a_0 \right) \, \nabla \phi_\text{N} \right],
\label{tempo}
\end{equation}

\noindent where \( \nu(y) \) represents a new interpolation function which
tends to \( y^{-1/2} \) far away from the strongest gravity regime. Our
approach is equivalent to their results, since the connection between \(
f(x) \) and \( \nu \) is given by

\begin{equation}
  f(x) = x^2 \left[ \nu(x^2) + 1 \right].
\label{tempo01}
\end{equation}

\noindent Comparing this last result with equation~\eqref{eq002} it
follows that \( \mu = \left( \nu + 1 \right)^{-1} \) and so, for \( \nu
\gg 1 \), then \( \mu = \nu^{-1} \).  For the specific value \( \nu(y)
= y^{-1/2} \) then

\begin{equation}
  f(x) = x + x^2.
\label{tempo02}
\end{equation}

\noindent The above two equations show that the function \( f \) is 
specifically a function of \( x^2 \) only, i.e. it depends exclusively on 
the Newtonian force \( x^2 \).

  In the following sections, we discuss several astrophysical systems
where the proposed gravity law~\eqref{star} can be tested
over a wide range of values of $x$, and explore the various predictions
which emerge. We perform comparisons mostly through expected scalings
related to velocity dispersions, masses and equilibrium radii, derived
from very general dimensional arguments. We consider mostly spherical
mass distributions \( M(r) \), making use of the validity of Newton's
theorems for spherical symmetry already discussed.

\subsection{Solar system consistency and an illustrative rotation curve of our Galaxy}
\label{solar}

As a first test of the proposed force law we compare the variations with
respect to Newtonian acceleration which would be introduced at solar
system scales, to the exquisitely measured upper limits on this quantity.

\begin{figure}
\begin{center}
  \includegraphics[scale=0.7]{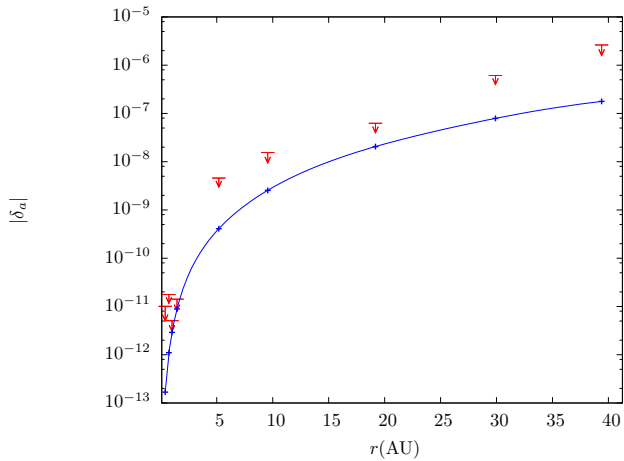}
\end{center}
\caption[Solar system]{The curve shows the fractional modifications \(
\delta a := \left( a - a_\text{Nt} \right)/ a_\text{Nt}\) our model
introduces when compared to purely Newtonian acceleration \( a_\text{Nt}
\). This is compared with the upper limits on the deviations from radial
Newtonian acceleration as reported by \citet{sereno06}.  The diagram
shows that the modifications introduced by our model are always within
the observed upper limits.  From left to right the crosses  represent
the calculated values \( \delta a \) at the distance of Mercury, Venus,
Earth, Mars, Jupiter, Saturn, Uranus, Neptune and Pluto respectively.}
\label{fig03}
\end{figure}

  For solar system scales with distances of between $0.1$ and $\sim 50$
AU and taking $M = 1 \, \textrm{M}_{\odot}$, we get $ 10^{2} \lesssim
x\lesssim 10^4$ (cf. Figure~\ref{fig07}).  Due to these large values,
we note immediately that the deviations from Newtonian dynamics
are negligible.  Figure~\ref{fig03} shows that our model lies
well within the observational upper limits for the departures from
Newtonian radial acceleration for the planets in our solar system,
as reported by \citet{sereno06}.  In consequence, under the proposed
force law, MONDian type dynamics have no relevance at solar system
scales. In light of our results, we argue in favour of studies by
\citet{anderson02,turyshev-toth,toth-turyshev}, that regard the anomaly of
the recession velocity of the Pioneer probes as an effect of the thermal
radiation of the spacecrafts \citep[see however, ][for interpretations
of the anomaly as a signature of MOND]{bekenstein06,milgrom-innerss}.

%%%%%%%%%%%%%%%%%%%%%%F I G U R E%%%%%%%%%%%%%%%%%%%%%%%%%%%%%%%%%%%%%%%%%
%%%%%%%%%%%%%%%%%%%%%%F I G U R E%%%%%%%%%%%%%%%%%%%%%%%%%%%%%%%%%%%%%%%%%

  As a particular example, we now turn to the rotation velocity curve
of our Galaxy.  The circular rotation velocity $V(r)$ associated to
equation \eqref{star} is given by

\begin{equation}
 V(r) = \left( a_0 l_\text{M} \, \frac{ 1 - x^4 }{ 1 - x^3 }\right)^{1/2}.
\label{eq09}
\end{equation}

\begin{figure}
\begin{center}
  \includegraphics[scale=0.7]{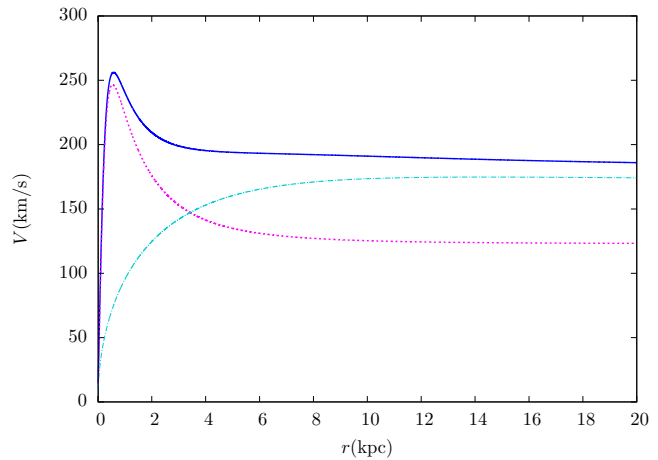}
\end{center}
\caption[Rotation]{The figure shows the rotation curve (continuous line) of 
our Galaxy using
the theory of gravity proposed in this article (cf. equation \eqref{star}).
We have assumed that only disc (dashed-dotted line) and bulge (dotted line)
contributions to the rotation curve are present.
}
\label{fig04}
\end{figure}

\noindent For the case of our Galaxy, the mass of the bulge and the mass of
the disc as a function of the radial distance are respectively given 
by \citep[see e.g.][]{allen} 

\begin{gather}
  m(r)\big|_\text{bulge} =  m  \, r^3 \, \left( a^2 + 
    r^2 \right)^{-3/2}, 				\label{eq10} \\
  m(r)\big|_\text{disc} = 2 \pi \Sigma_0 r_\star^2 \left[ 1 - \left(
    \frac{ r }{ r_\star } + 1  \right) e^{-r/r_\star} \right],
						\label{eq11}
\end{gather}

\noindent where \( m = 1.4 \times 10^{10} \, \mathrm{M}_\odot
\), \( a = 0.387 \, \mathrm{kpc} \), \( \Sigma_0 = 7 \times 10^8 \,
\mathrm{M}_\odot \, \mathrm{kpc}^{-2} \) and \( r_\star = 3.5 \,
\mathrm{kpc}\).  The effective mass at a distance \( r \) from the 
centre of our Galaxy is then given by

\begin{equation}
  M(r) = m(r) \big|_\text{bulge} + m(r)\big|_\text{disc}, 
\label{eq12}
\end{equation}

\noindent and so, by substitution of equation~\eqref{eq12}
into~\eqref{eq09} the theoretical rotation curve of our Galaxy, shown in
Figure~\ref{fig04}, is obtained. Given the validity of Newton's theorems
for spherical mass distributions in our case, the inner rotation curve of
Figure~\ref{fig04} is fully self consistent. This occurs since in that
region, the dynamics are completely dominated by the galactic bulge, which
has the same spherical symmetry as is assumed in the calculation. The
same holds to a large degree for the calculated rotation curve beyond
a few disc scale radii, where the mass distribution has essentially
converged. In the intermediate regime, the assumption of spherical
symmetry in the calculation necessarily introduces some error. However,
by analogy with the Newtonian case, where the rotation curve of an
exponentially decreasing infinitely thin disc deviates only about $20\%$
from that of the corresponding spherical mass distribution \citep[see
e.g][]{binney08}, this error is probably small.  The validation of
this analogy through a full 3D implementation of our model, lies beyond
the scope of this first presentation of our model.  In the light of this,
Figure~\ref{fig04} simply provides a graphical representation of the
fact that our \( f(x) \) was calibrated from our Galaxy's best fit \(
\mu(a/a_0) \) as given by \citet{binney-famaey}.

  This rotation curve represents well the observed features of
our Galaxy's rotation curve \citep[see e.g][]{GilK89}.  This was
actually built into our model, as the value of $n$ in the proposed force
law was calibrated from the numerical MOND interpolation function estimated
by \citet{binney-famaey}, which was obtained precisely from calibrations 
against the rotation curve of our Galaxy.

%%%%%%%%%%%%%%%%%%%%%%F I G U R E%%%%%%%%%%%%%%%%%%%%%%%%%%%%%%%%%%%%%%%%%
%%%%%%%%%%%%%%%%%%%%%%F I G U R E%%%%%%%%%%%%%%%%%%%%%%%%%%%%%%%%%%%%%%%%%

\section{Equilibrium radii of bound configurations}
\label{equilibrium}

If we think of an arbitrary astrophysical equilibrium structure with
total mass $M$, with characteristic equilibrium radius $r_\textrm{e}$
and internal velocity dispersion $\sigma$, we can estimate the relation
between the above three quantities by equating the kinematic pressure
to the mean gravitational force.  Note that in this section we use \(
M \) to denote the total mass of an extended astrophysical system.
Writing the kinematic pressure per unit mass as $C \rho \sigma^{2} /
M$, replacing $\rho$ for $M r_\textrm{e}^{-3}$ and equating this to the
proposed force law of equation~\eqref{star} we obtain

\begin{equation}
  C \tilde{\sigma}^{2} = \frac{R_\textrm{e}^{4}-1}{
    R_\textrm{e}^{4}-R_\textrm{e}},
\label{eqR}
\end{equation}

\noindent where we have introduced  a dimensionless velocity dispersion
$\tilde{\sigma} := \sigma /(G a_{0} M)^{1/4} $ and an equilibrium radius
$R_\textrm{e} := r_\textrm{e}/ l_\text{M}$ for the problem.  The constant
\( C \) is expected to be of order \( 1 \).

  We notice that for small values of $R_\textrm{e}$, corresponding to
large values of the variable $x$, i.e. the Newtonian regime, the right
hand side of equation~\eqref{eqR} reduces to $1/R_\textrm{e}$.  In this
case equation~\eqref{eqR} leads directly to the virial equilibrium
relation of Newtonian gravity $r_\textrm{e}= C^{1/2} \sigma /(G
\rho)^{1/2}$. At the opposite MONDian limit, with small values of
$x$ and so, large values of $R_\textrm{e}$, the right hand side of
equation~\eqref{eqR} tends to 1. In this limit, equation~\eqref{eqR}
yields the Faber-Jackson relation $C^{1/2} \sigma =(G a_{0} M)^{1/4} $.
Allowing for a proportionality between isotropic velocity dispersions in
pressure supported systems and rotation velocities in angular momentum
supported ones, this last relation could also be seen as the baryonic
Tully-Fisher relation which also has an index close to 4 \citep[see
e.g.][]{puech10}.  It is interesting that in terms
of the equilibrium gravitational radii of the proposed force law, the
galactic Faber-Jackson relation appears as the ``Jeans mass'' solution
for large values of $R_\textrm{e}$. Intermediate cases should be well
described by the appropriate points along the transition between the
two limits of equation~\eqref{eqR}.

  If we think of a gravitationally bound system with an associated
dispersion velocity \( \sigma \), the above results can also be
easily understood from the point of view of dimensional analysis.
Indeed, employing the same procedure we used in constructing
equation~\eqref{eq001}, the problem is now characterised by the dispersion
velocity \( \sigma \) of the system, the radius \( r \), the total mass \( M
\), Newton's constant of gravity \( G \) and Milgrom's constant \( a_0 \).
With these, Buckingham's theorem of dimensional analysis
demands that the velocity dispersion must have the following form

\begin{equation}
  \tilde{\sigma} = g(x),
\label{eq-dis}
\end{equation}
 
\noindent with  \( g(x) \) an arbitrary function to be determined.
A natural approximation often used, is to model observations over a
limited parameter range through power law representations.  As such, we
can explore the consequences of imposing a power law for the function \(
g(x) \), i.e.,

\begin{equation}
  \tilde{\sigma} \propto x^\alpha = l_\text{M}^\alpha \, r^{-\alpha}.
\label{eq15}
\end{equation}  

  By comparing equations ~\eqref{eqR} and ~\eqref{eq15}, we see
that equation~\eqref{eq15} is merely a power law approximation to the
full relation supplied by the proposed force law. One is lead to expect
that over any limited range of values of $x$, equation~\eqref{eqR} should
be accurately approximated by equation~\eqref{eq15}, with a suitable
choice of the parameter $\alpha$, which will more generally be a
function of $x$.

The strong prediction of equation~\eqref{star}, as seen in
equation~\eqref{eq15}, is that over any constrained range of values
of $x$, e.g. corresponding to any well defined class of astrophysical
objects, a relation of the type

\begin{equation}
\sigma=Cr^{\beta} M^{\gamma},
\label{sig}
\end{equation}

\noindent will always appear. Comparison with equation~\eqref{eq15}
yields directly $\beta=-\alpha$ and shows that the two power law indices
of equation~\eqref{sig} are not independent, but will obey the necessary
constraint 

\begin{equation}
\gamma=\frac{1}{4} - \frac{\beta}{2}.
\label{const}
\end{equation}

  Again, we see that the two limits explored above correspond
to $\alpha=1/2$ and $\alpha=0$, for the Newtonian and deep MOND
regimes respectively. It is clear that the constraint
of equation~\eqref{const}, obtained through dimensional analysis,
is satisfied in these two cases.  Note that this constraint can also be
derived through a power law approximation to the full force model of 
equation~\eqref{eqR}.  In what
follows it will become evident that this power law approximation and the
corresponding predicted constraint~\eqref{const} accurately reproduce the
empirical scalings observed in elliptical galaxies and the well studied
dwarf spheroidals of the local group.  We shall also present an exact
expression for $\alpha(x)$ at all scales as a strong testable prediction.

\subsection{The fundamental plane of elliptical galaxies}
\label{fundamental-plane}

The clearest correlations between the structural parameters of
elliptical galaxies are expressed by the fundamental plane relations,
first described by \citet{djorgovski,dressler}, as scaling relations
between the observed effective radius \( r \) of the Galaxy, the
central line of sight velocity dispersion $\sigma$, and the mean surface
brightness \( I \) within the effective radius \citep[see e.g.][for recent
observations]{bernardi,desroches}. Assuming a constant mass to light
ratio and using the fact that the mean surface brightness is given by the
luminosity $L$ within the effective radius, such that $I = L / 2 r^2$,
measurements over large samples, now in the thousands of galaxies, give a
relation identical to equation~\eqref{sig}. 

  In what follows we present a treatment for elliptical galaxies in
isolation.  Many such systems reside in galaxy clusters where external
field effects might have some relevance, e.g. in accounting for the escape
of high velocity stars \citep{wuto08}.  However, the well established
constancy of the basic scaling relations for ellipticals across different
density environments, implies that these external field effects are by far
not the physical causes driving the structure of the fundamental plane. As
such, we shall not consider such environmental effects further at this point.

  A broad agreement in the literature yields values close to
the ones reported by \citet{bernardi} of $\beta=-0.3356 \pm 0.023$ and
$\gamma= 0.503\pm 0.025$.  The first interesting point to note is that
the above measured values deviate for $\beta$, slightly but noticeably,
from the Newtonian expectations of $\beta=-1/2, \gamma=1/2$. This feature
is termed the tilt in the fundamental plane, and has traditionally been
explained in terms of mass to light ratios which vary as a function
of mass for elliptical galaxies. Although such explanations are not
unreasonable \citep[see e.g.][]{desroches}, we shall see that this tilt
in the fundamental plane can be accounted for, at constant mass to light
ratios, by the slight deviations from the Newtonian limit expected
from equation~\eqref{eqR} at the values of $R_\text{e}$ which correspond to
elliptical galaxies.

  As already seen from equation~\eqref{eqR} under the power law
approximation of equation~\eqref{sig}, we can compare the observed
values for the power law scalings of the fundamental plane of
equation~\eqref{sig} against the theoretical expectations of our
model through equation~\eqref{const}. For the observed value of
$\beta=-0.3356 \pm 0.023$, equation~\eqref{const} implies  $\gamma=
0.418 \pm 0.012$. This is marginally consistent with recent measurements
of $\gamma= 0.503 \pm 0.025$ at a two sigma level, under the hypothesis
of a constant average mass to light ratio for elliptical galaxies.
We therefore see that a power law approximation through a 
dimensional analysis approach to the problem furnishes the relation
between the indices \( \beta \) and \( \gamma \) in equation~\eqref{sig},
given by relation~\eqref{const}.  The actual value of both indices will
be obtained theoretically through the use of the full force model in
what follows.  It is noteworthy that practically all of the tilt in the
fundamental plane can be accounted for directly by the expectations of
equation~\eqref{const}.  The fullest explanation might well include a
certain degree of mass dependence for the average mass to light ratios,
as pointed out by various studies \citep[see e.g][]{desroches}.

In order to evaluate the parameter $\alpha$ we proceed by fitting a power
law function of the form~\eqref{eq15} to equation~\eqref{eqR}.  This can
be done locally to the smooth function~\eqref{eqR} by calculating the
tangent line to the function $\log\tilde {\sigma}(\log x)$ at a given
point $P_0(x_0, \tilde{\sigma_0})$, which satisfies the 
relation

\begin{equation}
  \log \tilde{\sigma} = \alpha(x_0) \log x + \log
       \tilde{\sigma_0} - \alpha(x_0) \log x_0.
\label{t08}
\end{equation}

\begin{figure}
\begin{center}
  \includegraphics[scale=0.7]{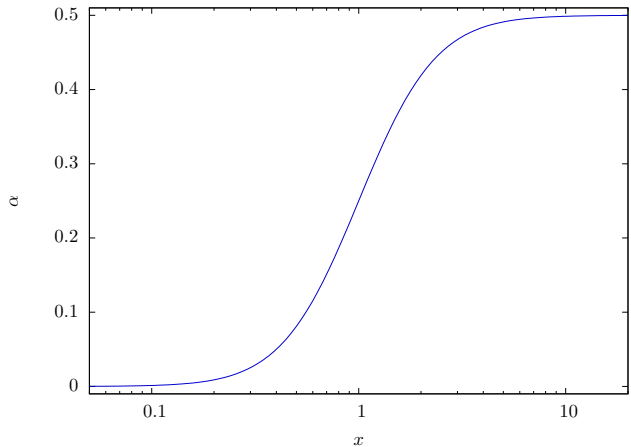}
\end{center}
\caption[Alphavsx]{The plot shows the variation of the power index
\( \alpha \) with respect to \( x \), in the general gravitational
equilibrium relation (see text) valid for all systems and at all scales.
For large values of the parameter \( x \), the Newtonian limit \(
\alpha = 1/2 \) is reached, and for small ones convergence to the MONDian
one \( \alpha = 0 \) is obtained.   As shown in Figure~\ref{fig07} many
astrophysical systems of interest fall in the intermediate regions about
\( x = 1 \).}
\label{fig05}
\end{figure}

\noindent The slope of the tangent line is given by

\begin{equation}
  \alpha(x) = \frac{\partial \log \tilde{\sigma}}
      {\partial \log x} = \frac{x^3}{2} \frac{3 + 2x + x^2}
      {(1 + x + x^2) (1 + x + x^2 + x^3)},
\label{t09}
\end{equation}

\noindent which simply means that the constant $\alpha$ depends
on the scale $x_0$ of the system as shown by Figure~\ref{fig05}.  Notice
that \( \alpha \to 1/2 \) for the Newtonian \( x \to \infty \) regime 
and \( \alpha \to 0 \) for the MONDian \( x \to 0 \) limit.

%%%%%%%%%%%%%%%%%%%%%%F I G U R E%%%%%%%%%%%%%%%%%%%%%%%%%%%%%%%%%%%%%%%%%
%%%%%%%%%%%%%%%%%%%%%%F I G U R E%%%%%%%%%%%%%%%%%%%%%%%%%%%%%%%%%%%%%%%%%

For any given class of astrophysical objects, the parameter $\alpha$ can now
be obtained by evaluating \(x \) and then solving for \( \alpha \) from equation~\eqref{t09}.

Note that typical values for elliptical galaxies with radii of
order a few kpc, and masses one order of magnitude around $10^{10}
\textrm{M}_{\odot}$ (see Figure~\ref{fig07}), we get \( x \sim 1 \),
which from equation~\eqref{t09} yields \( \alpha \sim 0.3 \). The tilt
in the fundamental plane is hence naturally explained by the force law
presented, requiring only minimum departures from constant average values
for the mass to light ratios of elliptical galaxies.

Recent studies spanning larger ranges of velocity dispersion and
mass values for elliptical galaxies \citep[e.g.][]{gargiulo,desroches}, 
have begun to identify slight trends in the power
indexes of the fundamental plane.  This is exactly what one would expect
from the developments presented here, as when one broadens the range of
physical parameters, the power law approximation to equation~\eqref{eqR}
ceases to be valid, and a warp will necessarily
develop. Being elliptical galaxies very close to the Newtonian limit of
$\alpha=1/2$, we can estimate the first order trend of this warp against
mass by considering that typical densities for elliptical galaxies tend to
drop as one goes to more massive systems, which would move the fit further
away from the Newtonian regime, leading to a gradual decrease in $\alpha$
with increasing total mass. By comparing with Figure~8 of \citet{gargiulo},
we see precisely this very trend, with the measured value
for their index \( a \) in the $\log r_\textrm{e} = a \log \sigma_{0}$
fit (notice that $a \propto \alpha^{-1}$) increasing as the limit
mass of the galaxies included in the sample increases.  Therefore, the
measured first order trends for the fundamental plane indexes with mass,
are seen to agree with the expectations of the model presented.

\subsection{Local dwarf spheroidal galaxies}

We now turn to the well studied dSph galaxies of the local group and their
scaling relations. Taking this time the first order terms for $R_\textrm{e}
\gg 1 $ in equation~\eqref{eqR}, we obtain

\begin{equation}
C \tilde{\sigma}^{2}  = 1 + \frac{1}{R^3_\textrm{e}}.
\end{equation}

\noindent Again we see that the first deviations from the deep MOND regime
occur after skipping the \( R_\textrm{e}^{-1} \) and \( R_\textrm{e}^{-2}
\) terms.  This implies that near the deep MOND regime, gravitational
physics remain largely unaltered.  If we now solve for 
$r_\textrm{e}$ from the above equation, we obtain

\begin{equation}
  r_\textrm{e}^{3}=\frac{(G M a_0 )^{2}}{a_{0}^3 \left[ C \sigma^{2}-(G
    M a_{0} )^{1/2}\right]}.
\label{Rdsph}
\end{equation}

  Through full integration of isothermal equilibrium density
profiles for the local dSph's under a force law equivalent to
what we propose here in the deep MOND regime, \citet{hernandez10}
showed that a suitable description of the problem can be found
without the need to invoke the presence of any dark matter. Still,
this might require the consistent inclusion of the disruptive
effects of galactic tides on the local dwarfs, as pointed out
by e.g.  \citet{Sanchez-Salcedo-Hernandez07,Angus08,McGaugh10}.
It is interesting to remark that \citet{hernandez10} also
showed that there is a very clear correlation between the mass
to light ratio resulting from the dynamical modelling in MOND type
prescriptions (or alternatively, the dark matter fraction for these
systems), and the relative ages of the stellar populations present,
as directly inferred from statistical studies of their resolved stellar
populations \citep[see e.g.][]{HernGil00,Dolphin02,Helmi08,Martin08,Tolstoy09}.
This correlation is natural in a scenario where only the stars present act
as a source of gravity, shining less as they age, but appears contrived
under the  dark matter hypothesis. Also, \citet{hernandez10}
showed the expected scaling on average between $\sigma^{4}$ and total
stellar mass for the objects in question, within observational errors,
expected from the deep MOND regime.  Introducing this scaling into
equation~\eqref{Rdsph} we obtain

\begin{equation}
r_\textrm{e} \propto \sigma^{2}.
\label{dsphSC}
\end{equation}

%%%%%%%%%%%%%%%%%%%%%%F I G U R E%%%%%%%%%%%%%%%%%%%%%%%%%%%%%%%%%%%%%%%%%
\begin{figure}
\begin{center}
  \includegraphics[scale=0.7]{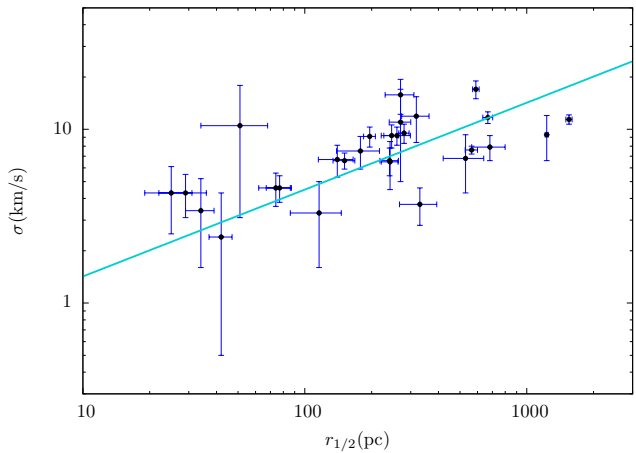}
\end{center}
\caption[Dwarfs]{Measured values of half light radii \( r_{1/2} \) and
internal velocity dispersions \( \sigma \) for the local dSph galaxies
with their associated one sigma uncertainties.  The solid line gives
a best fit $\sigma^2 \propto r_{1/2}$ relation, the exponent of
which results from the proposed force law.  Data from the compilation
of \citet{walker10}.  The point for the heavily distorted Sagittarius
dwarf shown at the right end has been excluded from the fit.}
\label{fig06}
\end{figure}
%%%%%%%%%%%%%%%%%%%%%%F I G U R E%%%%%%%%%%%%%%%%%%%%%%%%%%%%%%%%%%%%%%%%%

  We can now test the expectations of our proposed force law through
equation~\eqref{dsphSC}, by plotting the measured values of the velocity
dispersion for the local dSph galaxies, against their observed half-light
radii. We assume that our $r_\textrm{e}$ in equation~\eqref{dsphSC} will
be equal to some constant times the measured half light radius $r_{1/2}$
and take the compilation for these values found in \citet{walker10}
to plot Figure~\ref{fig06}. We see that in spite of the large errors
present, the expected scaling of $\sigma^2 \propto r_{1/2}$ given by the
line shown, clearly holds rather well. We have preferred this particular
plane to test our model, as velocity dispersions and half light radii
are the directly measured observables, not affected by the assumption
of rather uncertain mass to light ratios.

  Finally, we can try a first order extrapolation from the smallest
$R_\text{e} \ll1 $ systems, the  dSph galaxies, to the largest
astrophysical structures in that regime, galaxy clusters, by taking values
of $r=1 \textrm{kpc}$ and $\sigma=10 \, \textrm{km} \, \textrm{s}^{-1}
$ from Figure~\ref{fig06} and increasing the velocity dispersion by a
factor of 50. The expectation of equation~\eqref{dsphSC} then being radii
of \( 2.5 \times 10^{3} \) times larger. Indeed, \( 2.5 \, \textrm{Mpc} \)
and \( 500 \, \textrm{km} \, \textrm{s}^{-1} \) are representative values
for the structural parameters of large clusters of galaxies. Explaining
the dynamics of galaxy clusters in the pure MOND formalism requires some
unseen dark matter \citep[see e.g.][]{sanders03,angus10}, so we defer the
details of this problem to a latter study. Here we only point out that the
first order scaling of equation~\eqref{dsphSC}, extrapolated across \(
\sim 6 \) orders of magnitude in radius, is at least not significantly
in error.

Again, we see that having a well defined force law, rather than dealing
with a cumbersome MOND interpolation function, facilitates the physics
considerably. This allows a more direct tracing of the astrophysical
consequences, and yields more transparent predictions.  Notice that the
general results of the whole Section~\ref{equilibrium} are not dependent on
the details of \( f(x) \), and will hold for any such function within the
theoretical framework introduced.  We note however, the good agreement with
observations for the particular \( f(x) \) used.

\begin{figure}
\begin{center}
  \includegraphics[scale=0.7]{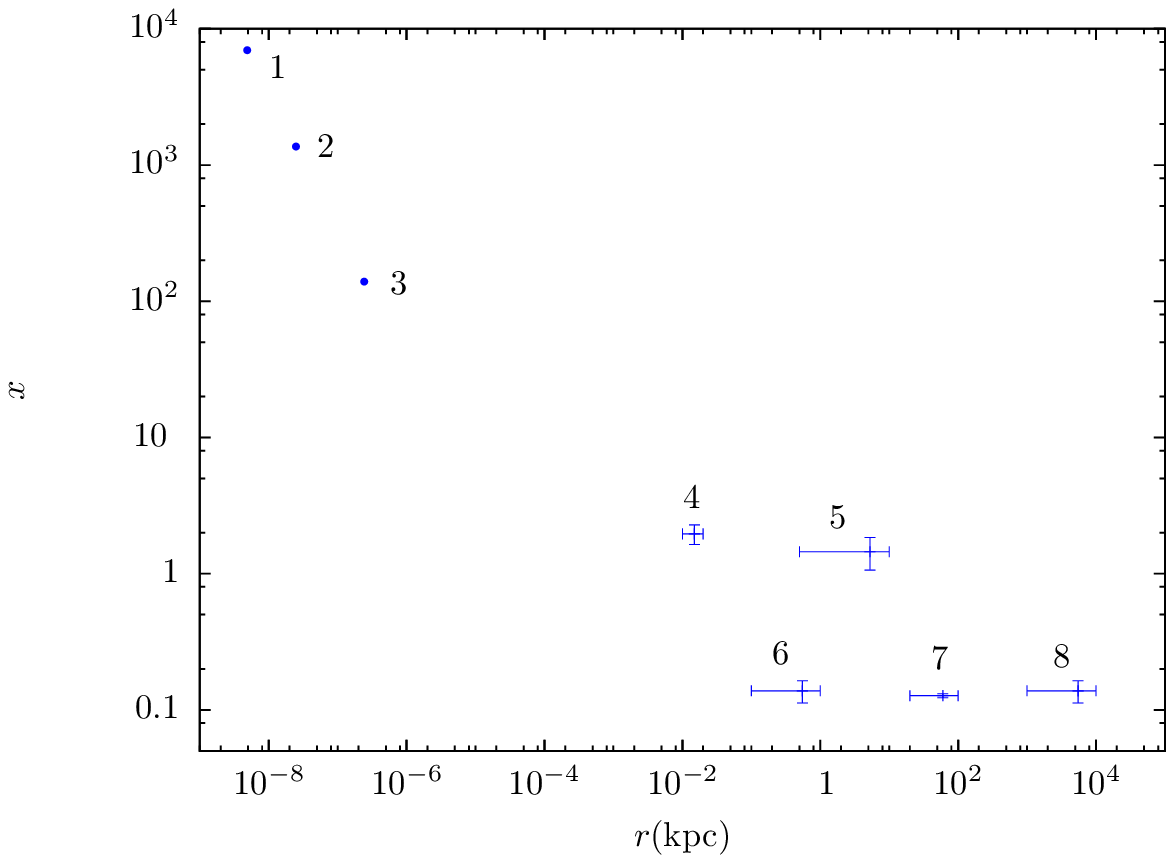}\\
  \includegraphics[scale=0.7]{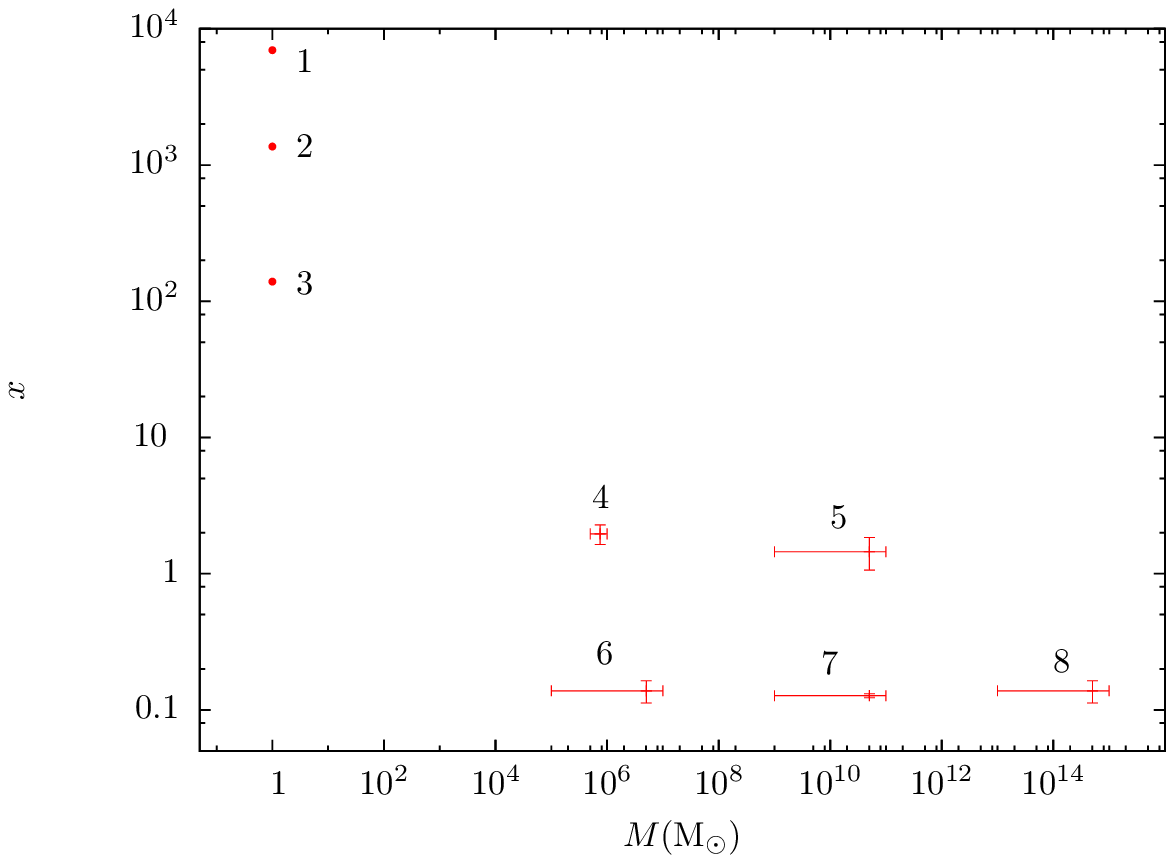}
\end{center}
\caption[AstroX]{The figure shows different values of \( x \propto
M^{1/2}/r \) for typical masses \( M \) and radii \( r \) of different
astrophysical systems: (\( 1 \))~Solar system at Earth's orbit.  (\(
2 \))~Solar system at Jupiter's orbit. (\( 3 \))~Solar system at the
Kuiper belt radius. (\( 4 \))~Globular clusters. (\( 5 \))~Elliptical
galaxies and bulges of spirals.  (\( 6 \))~Dwarf spheroidal galaxies. (\(
7 \))~Outer regions of spiral galaxies. (\( 8 \))~Galaxy clusters.}
\label{fig07}
\end{figure}

As a final summary of this section, we present Figure~\ref{fig07},
with typical values of the parameter $x$ for average masses and radii
for a range of astrophysical objects. This Figure, when compared to
Figure~\ref{fig05} clearly complements the results of the previous
sections.  We see that solar system dynamics will be utterly unaffected
by the modification proposed. Marginal modifications will appear
in the dynamics of globular clusters \citep[to within observational
uncertainties, see e.g.][]{hernandez10}. A slight tilt is to be expected
in the fundamental plane of elliptical galaxies and we see that the
external regions of spiral galaxies, galaxy clusters and local dwarf
spheroidal galaxies should appear as the most heavily ``dark matter
dominated'' systems.

%%%%%%%%%%%%%%%%%%%%%%F I G U R E%%%%%%%%%%%%%%%%%%%%%%%%%%%%%%%%%%%%%%%%%
%%%%%%%%%%%%%%%%%%%%%%F I G U R E%%%%%%%%%%%%%%%%%%%%%%%%%%%%%%%%%%%%%%%%%

\section{Concluding Remarks}
\label{discussion}

  We have shown in this article that if a modification of either
the dynamical or the gravitational part of Newton's gravity is to be
performed, a gravitational modification has a greater advantage for
physical clarity and ease of calculations.  Furthermore, it has become
clear that with the addition of Milgrom's acceleration constant \( a_0
\), all physical relations must be at least functions of the parameter
\( x \) defined in equation~\eqref{eq-dimensionless}, according to
Buckingham's theorem of dimensional analysis.   As examples, we have
shown that the acceleration felt by a particle on a centrally symmetric
gravitational field satisfies this condition.  Also, we show that the
velocity dispersion associated to a system in the linear approximation
gives rise to a generalised gravitational equilibrium relation which
converges to the well known results of the fundamental plane in the
description of elliptical galaxies, the Tully-Fisher relation for spiral
galaxies, the Faber-Jackson relation for elliptical galaxies, the Jeans'
stability criterion in the Newtonian limit and to corresponding observed
relations for dSph's galaxies.

  The inclusion of \( a_0 \) in the theory of gravity means that a
characteristic mass-length  \( l_\text{M} \) has been added.
This scale approximately indicates the transition between the MONDian and
Newtonian regimes.  By a suitable choice of model, this transition can
be built smoothly.  On very general grounds, a characteristic scale \(
l_\text{M} \) has been introduced into gravitation and so, gravity is
no longer scale invariant.

  We note also another testable prediction:  as precision increases
at solar system measurements the upper limits to departures from
Newtonian acceleration will converge to definite values, as implied
by Figure~\ref{fig03}.  

  The particular acceleration function we built is perhaps only an
approximation to a more general gravitational law.  The functions we
built, although logical and quite precise in principle, may just be good
candidates for the real gravitational theory and they should be thought
of as such: good approximations which fit a wide variety of available
astrophysical data to within errors.  However, it has to be through an
extension of the general theory of relativity that a more precise and
fundamental form is to be found.  This extended relativistic theory of
gravity must show that the PPN approximations converge in some limit to
the acceleration function built in this article.

  The Newtonian character of gravity has not been convincingly proven
across the astrophysical scales explored in this article and as such,
modifications to the theory are feasible. It is also important to note that
a very wide variety of modified gravity theories of the type developed
here, including MOND, are fundamentally falsifiable: a single \( a > a_0
\) system appearing as heavily dark matter dominated, would invalidate
them all as an alternative to the dark matter hypothesis.

\section{Acknowledgements}
\label{acknowledgements}
  This work was supported by a DGAPA-UNAM grant (PAPIIT IN116210-3).
The authors TB, XH \& SM  acknowledge economic support from CONACyT:
207529, 25006, 26344.  JCH acknowledges financial support from CTIC-UNAM.
The authors acknowledge T. Suarez \& L.A. Torres for help in various
aspects of the work presented here, and HongSheng Zhao pointing out the
connection between our work and the recent development of BIMOND and QMOND.
We thank the referee for valuable comments on the first version of the
article.  

%%%%%%%%%%%%%%%%
% BIBLIOGRAPHY %
%%%%%%%%%%%%%%%%
\bibliographystyle{mn2e}
\bibliography{my-mondian-world}

\label{lastpage}

\end{document}